# COMPARING SOFT COMPUTING TECHNIQUES FOR EARLY STAGE SOFTWARE DEVELOPMENT EFFORT ESTIMATIONS

Roheet Bhatnagar[1] and Mrinal Kanti Ghose[1]

[1]Department of Computer Science and Engineering, Sikkim Manipal Institute of Technology, Sikkim Manipal University, Majitar, Rangpo, East Sikkim, India

roheetbhatnagar@yahoo.com
mkghose2000@yahoo.com

## ABSTRACT

*Accurately estimating the software size, cost, effort and schedule is probably the biggest challenge facing software developers today. It has major implications for the management of software development because both the overestimates and underestimates have direct impact for causing damage to software companies. Lot of models have been proposed over the years by various researchers for carrying out effort estimations. Also some of the studies for early stage effort estimations suggest the importance of early estimations. New paradigms offer alternatives to estimate the software development effort, in particular the Computational Intelligence (CI) that exploits mechanisms of interaction between humans and processes domain knowledge with the intention of building intelligent systems (IS). Among IS, Artificial Neural Network and Fuzzy Logic are the two most popular soft computing techniques for software development effort estimation. In this paper neural network models and Mamdani FIS model have been used to predict the early stage effort estimations using the student dataset. It has been found that Mamdani FIS was able to predict the early stage efforts more efficiently in comparison to the neural network models based models.*

## 1. INTRODUCTION

Accurate estimation of software size, cost, effort and schedule is probably the biggest challenge facing software developers today. A typical estimation process involves generating a work breakdown structure (WBS), making assumptions, identifying dependencies, examining historical data, estimating each task and documenting the results [1]. Independent surveys carried out by Lederer [2] and Moløkken et al. [3] to evaluate the importance of effort estimation in software development, reported that 70-85% of the respondents agreed to the importance of effort estimation.. As software development has become an essential investment for many organizations, accurate software cost estimation models are needed to effectively predict, monitor, control and assess software development [4]. It has major implications for the management of software development because both the overestimates and underestimates have direct impact for causing damage to software companies. Since estimation accuracy is largely affected by modeling accuracy, finding good models for software estimation are now one of the most important objectives of the software engineering community [5]. New paradigms offer alternatives to estimate the software development effort, in particular the Computational Intelligence (CI) that exploits mechanisms of interaction between humans and processes domain knowledge with the intention of building intelligent systems (IS) [6]. Among IS, Artificial Neural Network and Fuzzy Logic are the two most popular soft computing techniques for software development effort estimation.

Since the last two decades, Artificial Neural Network (ANN) are being used extensively for predictions in diverse applications and the neural networks are recognized for their ability to produce reasonably accurate predictions in situations where complex relationships between inputs and outputs exist and where the input data is distorted by high noise levels [7]. Hughes [8], Wittig and Finnie [9][10] and Idri et al. [11] have employed neural network to predict the development effort on different data sets.

Many researchers have worked and proposed SCE models based on the Fuzzy Logic Techniques. Fei and Liu, [12] introduced the f-COCOMO model which applied Fuzzy Logic to the COCOMO model for software effort estimation. Kumar et al, [13] had applied fuzzy logic in Putnam's manpower buildup index (MBI) estimation model. Ryder [14] researched on the application of fuzzy logic to COCOMO and Function Points models. His result showed Fuzzy Logic is good at making effort estimations.

## 1.1. Early Stage Software Development Effort Estimations

Early stage effort estimations can be defined as making software development effort estimations at the initial stages more precisely the Design stage of SDLC. Carrying out effort estimations at the early stages is beneficial because the design stage prediction implies fewer overheads at the later stages of software development. Figure 1 below signifies that the total project effort comprises of the efforts (given in percentage) which goes into surpassing each of the individual phases. It is evident from the Figure 1 that most of the efforts (nearly 60 per cent) are spread over two initial phases of Analysis and Design. Hence if the accurate effort requirements can be predicted from the initial or early phases of the SDLC, then an efficient project development schedule can easily be prepared so as to complete the project well within the targeted time and budget constraints.

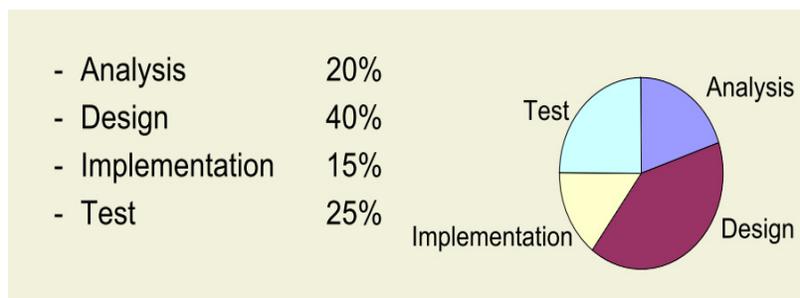

**Figure 1:** Effort distribution in the individual phases of SDLC
(Source: Peter Müller – Software Engineering, SS 2006)

The state of the art literature has revealed that not much work on estimating the effort required for software project development at the early stages in the Software Development Life Cycle (SDLC) has been done. Thus, this area still remains open to attract researchers to develop and propose new models for early stage effort estimation.

## 2. EXPERIMENTAL METHODOLOGY

For carrying out the effort prediction in the early stages of software development, precisely in the design phase of Software Development Life Cycle (SDLC), a student dataset was prepared based on the Entity Relationship Diagrams (ERDs) generated by the final year B.Tech. degree students of Computer Science & Engineering Department of Sikkim Manipal Institute of Technology, India, as part of their Major Project work spanning 16 weeks duration. Total Count of Entities (TCOE), Total Count of Attributes (TCOA), Total Count of Relationships (TCOR), Cumulative Grade Point Aggregate (CGPA) and Major Project final marks have been considered as explanatory variables in the dataset. The relevant data of students of different batches have been gathered. The final marks obtained by students in the Major Project are used to obtain the Recalculated Development Effort (RDE) in number of weeks (effort) of software development.

In a previous work [15] carried out by the authors of this paper, a comparison of different neural networks was carried out to predict the effort estimation at the early stages of software development. In the work the Development Time (DT) was obtained by applying various methods such as the Feed Forward Back Propagation Neural Network model, Cascaded Feed Forward Back Propagation Neural Network (CFFBPNN) model, Elman Back Propagation Neural Network (EBPNN) model, Layer Recurrent Neural Network (LRNN) model and Generalized Regression Neural Network (GRNN) model with the help of Neural Network toolbox of MATLAB R2007b software. The performances were then compared in terms of MMRE, Pred (0.25), BRE% etc. All these models were trained with first 31 inputs from the dataset and later the models were tested with 10 inputs from the same dataset.

In another work [16], Mamdani FIS from the Fuzzy logic toolbox of Matlab 7.0 was applied on the student dataset as given in Annexure II, Table 3, to evaluate the efficiency of the FIS in estimating the efforts in the early stages of SDLC. For experimentation from the dataset, the Total count of Entities (TCOE), Cumulative Grade Point Aggregate (CGPA) have been taken as two input variables and Redistributed Development Effort (RDE) as the output variable for preparing Mamdani FIS.

In the present paper a comparison of the performance of different neural network models with Mamdani FIS is done. For the experiments the same 'student dataset' was used and models were applied on to the dataset. A comparison of the MMRE values obtained from calculating the Redistributed Effort Estimations (RDE's) after employing the neural networks and fuzzy logic on the dataset was carried out to evaluate the efficiency of the better of the two in estimating effort estimation at the early stage of effort estimation.

### 2.1. Evaluation Criteria

There are many evaluation criteria to evaluate the accuracy of the software development effort in literature. The Mean Magnitude Relative error (MMRE) is a widely-accepted criterion in the literature and is based on the calculation of the magnitude relative error (MRE). Eq. (1) as below shows an equation for computing the MRE value that is used to assess the accuracies of the effort estimates.

$$MRE_i = \frac{|Actual\,Effort_i - \Pr edicted\,Effort_i|}{Actual\,Effort_i}$$

Eq. (1)

The MRE calculates each project in a dataset the MMRE aggregates the multiple MRE's. The model with the lowest MMRE is considered the best [4,17]. As shown in Eq. (2), the estimation accuracy of the MMRE is the mean of all the MREs among n software projects.

$$MMRE = \frac{1}{n}\sum_{j=1}^{n} MRE_j \qquad \text{Eq. (2)}$$

In this study, the MRE, and MMRE were adopted as the indicators of the accuracy of the established software effort estimation models since they are the ones most widely used in the literature, thereby rendering our results more comparable to those of other work.

## 3. RESULTS AND DISCUSSIONS

The values of MMRE are calculated for each of the neural networks and fuzzy logic are as shown in Annexure I, Table 2 and Annexure III, Table 4 respectively. The results obtained after comparing the RDE values are graphically shown in Figure 2 and their values are listed in Table 1.

Table 1 Comparison of different neural networks and Mamdani FIS based on MMRE values

| Models | MMRE(%) |
|---|---|
| FFBPNN | 12.96 |
| Cascaded FFBPNN | 13.59 |
| LRNN | 11.45 |
| **Mamdani FIS** | **3.89** |

**Figure 2:** Comparison of MMRE values of neural network and fuzzy logic

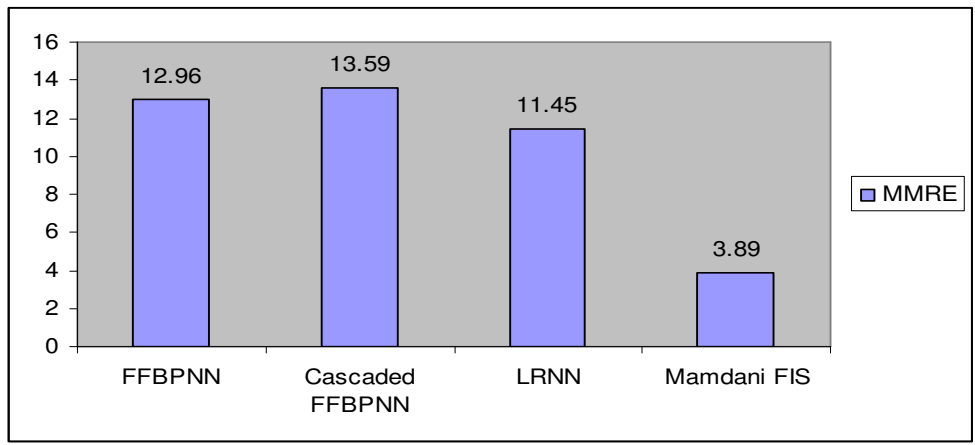

## 4. CONCLUSION

It is evident from the Figure 2 that the Linear Regression Neural network (LRNN) has the lowest value for MMRE among the other neural network models but when it is compared with fuzzy logic, it is observed that fuzzy logic outperforms neural network models as it has the lowest MMRE value. Thus, fuzzy logic is the best model for predicting early stage effort estimation. In an on-going work the efficacy of the models on real project dataset from the software industry will be established.

## **Annexure I**

Table 2: Development Effort as obtained by different neural network models along with their respective MRE and % MMRE values

| Serial No. | Actual RDE | RDE' using FFBPNN | MRE - FFBPNN | RDE' using Cascaded FBPNN | MRE – Cascaded FFBPNN | RDE' using LRNN | MRE - LRNN |
|---|---|---|---|---|---|---|---|
| 31 | 65 | 69.39 | 0.07 | 79.71 | 0.23 | 79.73 | 0.23 |
| 32 | 75 | 67.73 | 0.10 | 66.26 | 0.12 | 69.17 | 0.08 |
| 33 | 65 | 79.03 | 0.22 | 55.06 | 0.15 | 80 | 0.23 |
| 34 | 65 | 79.03 | 0.22 | 55.05 | 0.15 | 80 | 0.23 |
| 35 | 70 | 55 | 0.21 | 77.46 | 0.11 | 69.11 | 0.01 |
| 36 | 70 | 55.21 | 0.21 | 74.66 | 0.07 | 69.39 | 0.01 |
| 37 | 70 | 60.07 | 0.14 | 72.86 | 0.04 | 69.44 | 0.01 |
| 38 | 65 | 58.85 | 0.09 | 62.28 | 0.04 | 67.77 | 0.04 |
| 39 | 75 | 79.16 | 0.06 | 61.54 | 0.18 | 68.31 | 0.09 |
| 40 | 75 | 79.16 | 0.06 | 64.05 | 0.15 | 70.04 | 0.07 |
| 41 | 75 | 79.2 | 0.06 | 55.14 | 0.26 | 55.06 | 0.27 |
| **MRE VALUES** | | | **1.43** | | **1.49** | | **1.26** |
| **% MMRE VALUES** | | | **12.96** | | **13.59** | | **11.45** |

## Annexure II

**Table 3: ERD based Student Dataset**: TCOE :: Total Count of Entities; TCOA :: Total Count of Attributes; TCOR:: Total Count of Relationships; CGPA:: Parameter for academic excellence; RDE:: Redistributed Effort (Recalculated effort)

| Serial Number | TCOE | TCOA | TCOR | CGPA | RDE |
|---|---|---|---|---|---|
| 1 | 24 | 70 | 29 | 6.219 | 75 |
| 2 | 24 | 70 | 29 | 8.012 | 75 |
| 3 | 24 | 70 | 29 | 7.733 | 75 |
| 4 | 10 | 56 | 9 | 7.564 | 70 |
| 5 | 5 | 44 | 5 | 5.519 | 55 |
| 6 | 19 | 47 | 11 | 7.507 | 70 |
| 7 | 8 | 33 | 9 | 6.171 | 75 |
| 8 | 8 | 33 | 9 | 6.705 | 75 |
| 9 | 17 | 53 | 7 | 7.629 | 75 |
| 10 | 9 | 37 | 7 | 8.130 | 70 |
| 11 | 10 | 36 | 8 | 8.083 | 65 |
| 12 | 10 | 36 | 8 | 8.126 | 65 |
| 13 | 10 | 36 | 8 | 7.202 | 65 |
| 14 | 5 | 17 | 5 | 8.417 | 65 |
| 15 | 5 | 16 | 7 | 7.757 | 70 |
| 16 | 4 | 26 | 4 | 7.431 | 70 |
| 17 | 4 | 26 | 4 | 7.121 | 70 |
| 18 | 4 | 26 | 4 | 7.660 | 70 |
| 19 | 7 | 34 | 6 | 8.017 | 75 |
| 20 | 7 | 34 | 6 | 9.076 | 75 |
| 21 | 7 | 27 | 5 | 7.550 | 70 |
| 22 | 6 | 37 | 5 | 6.583 | 65 |
| 23 | 6 | 27 | 12 | 7.276 | 65 |
| 24 | 6 | 27 | 12 | 8.124 | 65 |
| 25 | 5 | 26 | 4 | 6.530 | 75 |
| 26 | 5 | 26 | 4 | 6.685 | 70 |
| 27 | 6 | 28 | 6 | 7.843 | 65 |
| 28 | 7 | 38 | 9 | 9.160 | 70 |
| 29 | 7 | 38 | 9 | 8.617 | 75 |
| 30 | 6 | 18 | 3 | 8.719 | 80 |
| 31 | 4 | 22 | 3 | 8.860 | 65 |
| 32 | 5 | 18 | 5 | 7.664 | 75 |
| 33 | 16 | 85 | 15 | 6.795 | 65 |
| 34 | 16 | 85 | 15 | 6.757 | 65 |
| 35 | 9 | 36 | 9 | 6.207 | 70 |
| 36 | 9 | 36 | 9 | 6.636 | 70 |
| 37 | 9 | 36 | 9 | 6.790 | 70 |
| 38 | 8 | 24 | 7 | 8.095 | 65 |
| 39 | 20 | 115 | 22 | 7.990 | 75 |
| 40 | 20 | 115 | 22 | 8.095 | 75 |
| 41 | 15 | 60 | 9 | 6.340 | 75 |

# Annexure III

Table 4: RDE using Mamdani FIS and corresponding MRE values

| Serial Number | TCOE | CGPA | RDE | RDE using Mamdani FIS | MRE |
|---|---|---|---|---|---|
| 1 | 24 | 6.219 | 75 | 75 | 0.000 |
| 2 | 24 | 8.012 | 75 | 75 | 0.000 |
| 3 | 24 | 7.733 | 75 | 75 | 0.000 |
| 4 | 10 | 7.564 | 70 | 75 | 0.071 |
| 5 | 5 | 5.519 | 55 | 64.3 | 0.169 |
| 6 | 19 | 7.507 | 70 | 75 | 0.071 |
| 7 | 8 | 6.171 | 75 | 65 | 0.133 |
| 8 | 8 | 6.705 | 75 | 65 | 0.133 |
| 9 | 17 | 7.629 | 75 | 75 | 0.000 |
| 10 | 9 | 8.13 | 70 | 75 | 0.071 |
| 11 | 10 | 8.083 | 65 | 75 | 0.154 |
| 12 | 10 | 8.126 | 65 | 75 | 0.154 |
| 13 | 10 | 7.202 | 65 | 75 | 0.154 |
| 14 | 5 | 8.417 | 65 | 71 | 0.092 |
| 15 | 5 | 7.757 | 70 | 71 | 0.014 |
| 16 | 4 | 7.431 | 70 | 70 | 0.000 |
| 17 | 4 | 7.121 | 70 | 70 | 0.000 |
| 18 | 4 | 7.66 | 70 | 70 | 0.000 |
| 19 | 7 | 8.017 | 75 | 73.4 | 0.021 |
| 20 | 7 | 9.076 | 75 | 72.8 | 0.029 |
| 21 | 7 | 7.55 | 70 | 73.2 | 0.046 |
| 22 | 6 | 6.583 | 65 | 64.4 | 0.009 |
| 23 | 6 | 7.276 | 65 | 71.3 | 0.097 |
| 24 | 6 | 8.124 | 65 | 72.1 | 0.109 |
| 25 | 5 | 6.53 | 75 | 64.4 | 0.141 |
| 26 | 5 | 6.685 | 70 | 64.5 | 0.079 |
| 27 | 6 | 7.843 | 65 | 72.1 | 0.109 |
| 28 | 7 | 9.16 | 70 | 72.7 | 0.039 |
| 29 | 7 | 8.617 | 75 | 73.3 | 0.023 |
| 30 | 6 | 8.719 | 80 | 71.9 | 0.101 |
| 31 | 4 | 8.86 | 65 | 70 | 0.077 |
| 32 | 5 | 7.664 | 75 | 71 | 0.053 |
| 33 | 16 | 6.795 | 65 | 70 | 0.077 |
| 34 | 16 | 6.757 | 65 | 70.4 | 0.083 |
| 35 | 9 | 6.207 | 70 | 67.1 | 0.041 |
| 36 | 9 | 6.636 | 70 | 68.6 | 0.020 |
| 37 | 9 | 6.79 | 70 | 70 | 0.000 |
| 38 | 8 | 8.095 | 65 | 75 | 0.154 |
| 39 | 20 | 7.99 | 75 | 75 | 0.000 |
| 40 | 20 | 8.095 | 75 | 75 | 0.000 |
| 41 | 15 | 6.34 | 75 | 71 | 0.053 |

## Authors

**Dr. Roheet Bhatnagar** received his B.Tech. in Computer Science and Engineering and M.Tech. in Remote Sensing from Birla Institute of Technology, Mesra, Ranchi, India in 1996 and 2004 respectively and PhD in Computer Science & Engineering from Sikkim Manipal University in 2011. He is having more than 14 years of varied experience in the software industries and academics. He had worked in multinationals viz; Xerox Modicorp Ltd., Samsung SDS India Pvt. Ltd. and USHA Soft (a software subsidiary of USHA Martin Ltd.) in Gurgaon from 1997 till 2003 just after his graduation. During his stint in the industry he had a good exposure to software 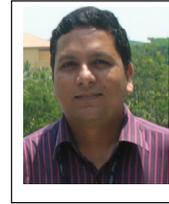 development executing many projects with different roles and responsibilities. He joined Department of Remote Sensing at BIT Mesra, Ranchi in the year 2003 and worked as Assistant Professor till 2008. He joined Sikkim Manipal Institute of Technology (SMIT) - a constituent college of Sikkim Manipal University (SMU) in 2008 and is presently serving as Associate Professor in the Department of Computer Science and Engineering. He has a number of publications in indexed international journals and national and international conferences. He is a life member of professional societies like Indian Society of Remote Sensing (ISRS), Indian Society of Technical Education (ISTE), and International Association of Engineers (IAENG). His current areas of interest are, soft computing, fuzzy and neural networks, database management systems, data mining and knowledge discovery, Remote Sensing and Geographical Information Systems (RS-GIS), and software engineering.
He can be reached at roheetbhatnagar@yahoo.com and roheet.bhatnagar@gmail.com

**Prof. (Dr) Mrinal Kanti Ghose** was born on 1st March 1952. He is a PhD and specializes in Software Engineering, Image Processing, Remote Sensing & GIS. His other Area of research are Artificial Intelligence, Data Mining, Simulation & Modeling, Optimization & Genetic Algorithms.

Currently Prof. Ghose is working as Dean (R&D), SMIT and Professor & Head, Department of Computer Science & Engineering at Sikkim Manipal Institute of Technology, 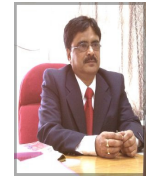

Sikkim, India. He is having vast experience of 32 years in teaching and research. During his career he has been associated with many prestigious universities and organizations. He had worked at Regional Engg. College ( NIT ), Silchar (1979 – 1981), Assam Central University, Silchar as COE and HOD of Computer Science Department (1997-2000). He was associated with Vikram Sarabhai Space Centre / ISRO, Thiruvananthapuram from 1981-1994 & Regional Remote Sensing Service Center / ISRO , Kharagpur from 1995 – 1996 and from 2000-2006. He was an Adjunct Professor, Reliability Engg Centre, IEM, IIT Kharagpur, from 2000 – 2005. He has more than 95 research publications in reputed National/International Journals and Conferences. He has written a number of Technical reports and co-authored a couple of books. He has organized a number of Conferences, Workshops and Seminars. He has guided a number of Master level students and guiding a number of PhD students. He has also worked on a number of consultancy projects.